\documentclass[useAMS,usenatbib]{mn2e}

\usepackage{graphicx}
\usepackage{multirow}
\usepackage{amsmath}
\usepackage[T1]{fontenc}
\usepackage{ae,aecompl}

\title[Champagne Flutes and Brandy Snifters]
{Champagne Flutes and Brandy Snifters: Modelling Protostellar Outflow--Cloud Chemical Interfaces}
\author[R.P.Rollins, J.M.C.Rawlings, D.A.Williams and 
M.P.Redman]{R. P. Rollins$^1$\thanks{E--mail: 
rpr@star.ucl.ac.uk}, J. M. C. Rawlings$^1$, D. A. 
Williams$^1$ and M. P. Redman$^2$\\ $^1$Department of Physics 
and Astronomy, University College London, Gower Street, London 
WC1E 6BT, United Kingdom \\ $^2$Centre for Astronomy, School 
of Physics, National University of Ireland Galway, University Road, Galway, 
Ireland}

\begin{document}

\maketitle

\begin{abstract}

\noindent A rich variety of molecular species has now been observed towards hot cores in star forming regions and in the interstellar medium. An increasing body of evidence from millimetre interferometers suggests that many of these form at the interfaces between protostellar outflows and their natal molecular clouds. However, current models have remained unable to explain the origin of the observational bias towards wide--angled ``brandy snifter'' shaped outflows over narrower ``champagne flute'' shapes in carbon monoxide imaging. Furthermore, these wide--angled systems exhibit unusually high abundances of the molecular ion HCO$^+$. We present results from a chemo--dynamic model of such regions where a rich chemistry arises naturally as a result of turbulent mixing between cold, dense molecular gas and the hot, ionized outflow material. The injecta drives a rich and rapid ion--neutral chemistry in qualitative and quantitative agreement with the observations. The observational bias towards wide--angled outflows is explained naturally by the geometry--dependent ion injection rate causing rapid dissociation of CO in the younger systems. 

\end{abstract}


\section{Introduction}
\label{sec:intro}

Our understanding of the formation and evolution of protostars has evolved into a highly complex and three dimensional model. Since the one dimensional inside out collapse theory of \cite{S:1977}, we now observe many distinct features such as disks, outflows, jets, HII regions and turbulence to be important in governing these systems. To develop an understanding of such features is to further our understanding of the star formation process on all scales. Of particular interest are the molecular outflows observed in low mass protostars. They have the potential to liberate accreting material of angular momentum as well as impacting dynamically on the surrounding molecular core. This creates a turbulent mixing zone which can harbour a rich and unique chemistry. Spectroscopic observations now show that many molecules are harboured in the outflow--cloud interface of these objects, particularly carbon monoxide \citep{Yea:2010}, water (\citealt{Kea:2010}, \citealt{Nea:2010}, \citealt{vKea:2010}, \citealt{Hea:2012}) and HCO$^+$ \citep{Hea:2013}. \cite{CVWB:2006} have also identified methanol, deuterated water, OCS and H$_2$CS in the high--mass analogue CepA--East. An understanding of the dynamics and chemistry of such systems is therefore clearly important for a complete understanding of the star formation process. 

Observations of low mass protostellar outflows exhibit two particularly interesting chemical features that can help diagnose the physical conditions. The first is an apparent observational preference towards wide angled cavities mapped in carbon monoxide within a few thousand astronomical units of the central source (for example, B5 IRS1, \citealt{LVX:1996}, HH 46/47, \citealt{Aea:2013}, \citealt{Rea:2013}). We liken the observed morphologies to ``brandy snifters'' -- short hour--glass shapes with wide opening angles at the base and a given height at which the width becomes maximal. By analogy, there are few observed molecular outflows shaped like ``champagne flutes'' -- narrow opening angles at the base, tall and thin all the way up. The apparent absence of ``champagne flute'' shaped outflows from observations is an issue that has not been adequately addressed by current models. Many attempts have been made to understand the carbon monoxide morphologies in terms of the dynamics of the outflow--cloud system. Initially, \cite{GGB:1996} presented a model for a precessing cavity and argued that the observed CO maps were a combined result of those dynamics with radiative transfer effects. More recently, \cite{LWM:2013} modelled the cavity shape as a result of turbulent entrainment and again produced synthetic observations using radiative transfer for the high--mass source G240.31+0.07 to be compared with observations by \cite{Qea:2009}. Their results suggest that ``champagne flute'' shaped outflows should in fact be observable. While both methods can produce qualitatively accurate synthetic observations for individual objects, they both make the simplifying assumption of a fixed CO abundance with polar angle along the interface. No attempt is made to consider three dimensional variations in the carbon monoxide chemistry throughout the interface as the origin of the observed morphology. \cite{Vea:2012} attempted such a model; a static cavity with heating by ultraviolet (UV) radiation and local C--shocks driving the chemistry. However, by considering only fits to observed objects those authors were unable to address a potential chemical origin of the morphological bias.

Secondly, abundances of HCO$^+$ in the outflows are observed to be significantly higher than estimated from modelling of dark clouds (e.g. L1527, \citealt{Hea:1998}). In a previous attempt to model HCO$^+$ formation, \cite{RTW:2000} proposed that the high abundance was a result of photoionized carbon reacting with water liberated from dust grain surfaces by shocks. Follow--up work showed that chemical variations along the cavity wall could reproduce HCO$^+$ observations \citep{Rea:2004}. However, the increased abundance is only temporary as the dissociative recombination of HCO$^+$ is rapid (10--100 years) under typical conditions, especially temperatures below $100~{\rm K}$. The temporary nature of its enhancement would suggest that in a decollimating outflow, it would only be present near the central source, and requires that cavities steadily grow over time, otherwise the enhancement disappears from the system on a 100 year timescale.

Photochemistry is certainly not the only possible process driving the chemistry in such systems. In this paper we develop and investigate a model for the chemistry in these cavity walls driven by hot ions from the protostellar ejecta mixing with molecular gas, with the aim of understanding the observations of CO and HCO$^+$ summarised above.
In particular, we speculate that the degree of interaction between the outflow and the 
interface (and hence the dynamically-induced molecular enrichment) is a function of 
the impact angle between them. This could help explain the observed bias towards certain morphologies.

We describe in Section \ref{sec:cavity} a parametric model for the dynamics of outflows to quantify the ion injection rate and also the key dynamical timescales. In Section \ref{sec:chemistry}, we present the chemistry used in our model, discussing assumptions on the initial conditions and chemical timescales. The results of our combined chemo--dynamical model are presented in Section \ref{sec:results} and discussion with respect to observations and free parameters in Section \ref{sec:discussion}. Our final conclusions are presented in Section \ref{sec:conclusions} along with brief proposals for further work.

\section{Cavity Injection Model}
\label{sec:cavity}

\begin{figure}
\centering
\includegraphics[scale=0.4]{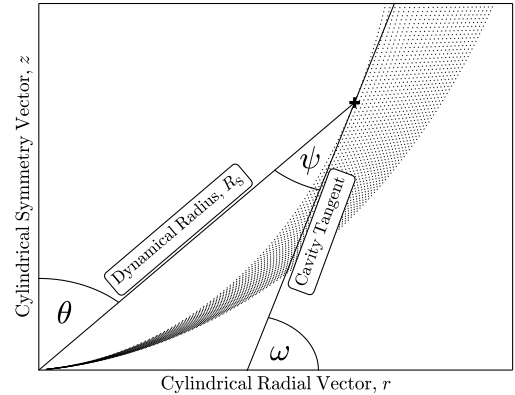}
\caption{Cross section of the model geometry for an example cavity--cloud interface. The shaded area shows the interface with its thickness exaggerated to be 50$\%$ of the dynamical radius for clarity. The tangent to the cavity wall is shown for a given point along with the corresponding polar angle, $\theta$, tangent gradient, $\omega$, and turbulent mixing angle, $\psi$.}
\label{fig:diagram}
\end{figure}

In order to quantify the rate at which material may be injected into the outflow--cloud interface, we must first be able to characterise its dynamical properties, critically the shape and velocity and how these may vary as a function of age. A number of possible origins for the structure of such cavity walls have been discussed, most notably magnetohydrodynamic (\citealt{Bea:2011}, \citealt{BTP:2014}) and turbulent entrainment (\citealt{Cea:2006}, \citealt{LWM:2013}). However, for the sake of our model, it is not important to understand the physical origin of the cavities but rather to be able to parametrise their typical observed shapes. An elegant example of this approach was presented by \cite{CRW:2008} for the cavity wall of Barnard 5 IRS 1. Their dynamical model is of an outflow coming from the central source with constant total mass loss rate ($\dot{M}_{\rm 0}$) and velocity ($v_{\rm 0}$) and steadily decollimating at a constant angular rate ($\epsilon$, the rate at which the outflow opening angle increases in radians per unit time). The outflow impacts on an isothermal sphere of density $\rho(R) = \kappa R^{-2}$ where $R$ is the radius and $\kappa$ is a scaling coefficient treated as a free parameter. Those authors were able to reproduce the observed shape and inferred dynamical properties whose parameters are summarised in Table \ref{table:outflowparams}. We choose to adopt this model since it not only gives the parametrised shape of a typical wide--angled source, but also defines cavity shapes as a function of dynamical age; younger ages correspond to narrower cavities, and vice versa. This allows us to directly test the chemical origin of the observational bias towards wider opening angles.

\begin{table}
\centering
\caption{Physical parameters for the cavity in B5 IRS 1 \citep{CRW:2008} used for our dynamical model.}
\label{table:outflowparams}
\begin{tabular}{ | c | c | }
  \hline
  \multicolumn{2}{|c|}{Cavity Parameters}  \\
  \hline
  Scale Length, $r_{\rm 0}$ & 350 ~{\rm au} \\
  Mass Loss Rate, $\dot{M}_{\rm 0}$ & $2.9\times10^{-6}~{\rm M_\odot~yr^{-1}}$ \\
  Outflow Velocity, $v_{\rm 0}$ & $100~{\rm km~s^{-1}}$ \\
  Decollimation Rate, $\epsilon$ & $1.05\times10^{-4}~{\rm rad~yr^{-1}}$ \\
  Isothermal Coefficient, $\kappa$ & $6.7\times10^{14}~{\rm g~cm^{-1}}$ \\
  \hline
\end{tabular}
\end{table}

Given the radial shape equation for the cavity, $R_{\rm S}(\theta,t_{\rm dyn})$, as a function of polar angle, $\theta$, and cavity age, $t_{\rm dyn}$, we assume an interface forms between the cavity and gas envelope with thickness equal to 10$\%$ of the dynamical radius; a value suggested by observations (e.g. \citealt{LVX:1996}). Here, the radially outflowing material impacts on the interface at an angle $\psi$ (as derived in Appendix \ref{app:psi} and shown in Figure \ref{fig:diagram}) given by the equation:

\begin{equation}
\label{eq:psimix}
\psi = \mathrm{arctan}\left(\frac{R_{\rm S}' - R_{\rm S}\tan\theta}{R_{\rm S}'\tan\theta + R_{\rm S}}\right) + \theta - \frac{\pi}{2}
\end{equation}

\noindent where $R_{\rm S}'$ is the derivative ${{\rm d}R_{\rm S}}/{{\rm d}\theta}$. This angle is then used with the other parameters to characterise the rate at which material from the outflow mixes turbulently at each point into the interface, as derived in Appendix \ref{app:ndot}:

\begin{equation}
\label{eq:injection}
\dot{n}_{\rm inj} = \frac{5\dot{M}_{\rm 0}}{\pi m_{\rm H} [R_{\rm s}(\theta)]^3} \frac{\sin\psi}{1-\cos[\epsilon(t_{\rm dyn}-\frac{R_{\rm S}(\theta)}{v_{\rm 0}})]} \, {\rm cm}^{-3} \, {\rm s}^{-1}
\end{equation}

\noindent where $m_{\rm H}$ is the atomic mass of hydrogen. The factor sin$\psi$ is a nominal mixing efficiency chosen on the assumption that only the component of the outflow normal to the cavity wall will be able to mix turbulently into the interface. A key prediction of this work is that this mixing angle is the key parameter in determining the local injection rate at each point in the interface. We expect that only those regions where $\psi$ takes a large value are able to form significant abundances of CO and HCO$^+$ due to the enhanced chemical interaction between the gas from the envelope and the injected material. The choice of function for this mixing efficiency and its impact on the model are considered in Section \ref{sec:discussion}.

\begin{figure}
\centering
\includegraphics[scale=0.4]{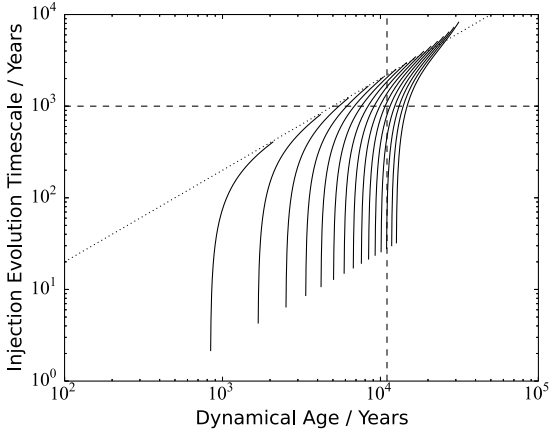}
\caption{Iso--angle tracks for the timescale on which the injection rate (Equation \ref{eq:injection}) changes. Each solid line represents one fixed value of $\theta$ between 4 and 60 degrees and shows the timescale evolving with the dynamical age of the growing cavity at that polar angle. The dotted line $\tau_{\rm inj}=t_{\rm dyn}/5$ is an approximate asymptote and shows that this timescale approaches approximately 20$\%$ of the dynamical age for all angles. The dashed vertical line shows the dynamical age of B5 IRS1 ($11~{\rm kyrs}$) inferred by \protect\cite{CRW:2008}. The dashed horizontal line demonstrates that for most angles, by a cavity age of $11~{\rm kyrs}$, the timescale on which the injection rate changes is more than $1~{\rm kyr}$. As discussed in Section \ref{sec:chemistry}, the chemistry of the interface reaches a quasi--equilibrium within this time. $1~{\rm kyr}$ is therefore a suitable chemical integration time for our decoupled chemical model at all but the angles close to the opening angle. For younger cavities, shorter chemical timescales will be suitable.}
\label{fig:timescale}
\end{figure}

There are three key timescales associated with such a system. One is the timescale on which the cavity grows, which is simply $R_{\rm S}/({{\rm d}R_{\rm S}}/{{\rm d}t_{\rm dyn}})$. We note that this tends towards approximately one fifth of dynamical age of the cavity on all angular scales. The second is the timescale on which the interface density increases and its composition becomes dominated by the material injected from the outflow into the interface, given by $n_{\rm 0} / \dot{n}_{\rm inj}$ where $n_{\rm 0}$ is the initial particle density in the interface. While the interface density is a free parameter of our model, typical values for the density and cavity age give a timescale many orders of magnitude longer than the dynamical age, so that the composition of the interface is never dominated by the injected material. However, even if the composition of the cavity is not dominated by hot ions from the outflow, their presence can still drive the dominant chemical processes.

Instead, it is the timescale on which the injection rate varies, $\tau_{\rm inj}=\dot{n}_{\rm inj}/({{\rm d}\dot{n}_{\rm inj}}/{{\rm d}t_{\rm dyn}})$, that is important to our problem and visualised in Figure \ref{fig:timescale}. Again, we see that it approaches approximately a fifth of the dynamical age of the cavity for all angles. As a result, any chemistry driven purely by the injecta that can reach equilibrium on timescales an order of magnitude shorter than the dynamical age can be modelled independently of the dynamics since the injection rate is effectively constant. However, if the chemical timescales are significantly longer than the dynamical age then it is necessary to integrate the dynamics and chemistry coupled in time. As we shall further argue in Section \ref{sec:chemistry}, the much simpler first situation holds for our model, with chemical equilibrium times of order $1~{\rm kyr}$ being typical. 

\section{Chemical Model}
\label{sec:chemistry}

Given that the outflowing material interacts dynamically with the envelope to form the observed interface, we now wish to model the chemical interactions between the two gases as they mix turbulently. Our chemical model borrows heavily from the work of \cite{RTW:2000} in modelling outflow interface chemistry. In their work, HCO$^+$ is formed as a product of the reaction between C$^+$ and water:

\[ {\rm C^+ + H_2O \rightarrow HCO^+ + H.}\] 

\noindent The available C$^+$ was presumed to have been formed from the photoionization of atomic carbon by UV radiation given off in the bow shock associated with the expansion of the cavity. At the same time, the shock was assumed to liberate the icy mantles of dust grains, giving an initial gas phase chemical composition equal to typical dark cloud dust mantle compositions, including the necessary water for processing. Indeed, \cite{BP:1997}  observed a rich array of species produced in the shocked regions of L1157, including some believed to be processed from interstellar ices. More specifically, \cite{Nea:2010} see emission from water that is spatially correlated with the outflow shock of L1157. While this reaction pathway could generate high fractional abundances of HCO$^+$ ($>$10$^{-7}$), it was only a temporary feature as the dissociative recombination reaction:

\[ {\rm HCO^+ + e^- \rightarrow CO + H}\] 

\noindent was rapid at typical conditions, and especially below $100~{\rm K}$ due to its strong temperature dependence. Hence, the high abundances in \cite{RTW:2000} were only realised for 10--100 years.

Our chemical model includes the species shown in Table \ref{table:species}. We consider similar initial conditions to those adopted by \cite{RTW:2000}, with the initial gas--phase fractional chemical abundances, $Y_{\rm 0}$, coming from observationally inferred values for the ice composition of dark clouds. These values are given, as a function of the total elemental abundances, $X$, in Table \ref{table:initial}. The gas is thus assumed to
be fully molecular, apart from an atomic hydrogen abundance of $1~{\rm cm^{-3}}$, 
established by the balance between cosmic ray-induced destruction and grain
surface formation of H$_2$.

\begin{table}
\centering
\caption{List of chemical species included in our model.}
\label{table:species}
\begin{tabular}{c}
\hline
Chemical Species \\ 
\hline
H, H$^+$, H$_2$, H$_2$$^+$, H$_3$$^+$, H$^-$ \\
He, He$^+$, Na, Na$^+$, e$^-$ \\
C, C$^+$, CH, CH$^+$, CH$_2$, CH$_2$$^+$ \\
CH$_3$, CH$_3$$^+$, CH$_4$$^+$, CH$_4$, CH$_5$$^+$ \\
O, O$^+$, O$_2$, O$_2$$^+$, OH, OH$^+$, H$_2$O, H$_2$O$^+$, H$_3$O$^+$ \\
CO, CO$^+$, HCO, HCO$^+$, H$_2$CO, H$_2$CO$^+$ \\
\hline
\end{tabular}
\end{table}
\null

\begin{table}
\centering
\caption[table]{Total elemental abundances and initial fractional abundances of 
gas-phase species, relative to hydrogen nucleons.}
\label{table:initial}
\begin{tabular}{|l|l|l|}
\hline
 Parameter & Definition & Value \\
\hline
 $X$(He) & He/H & $0.0763$  \\
 $X$(C)  & C/H  & $2.0\times10^{-4}$ \\
 $X$(O)  & O/H  & $3.02\times10^{-4}$ \\
 $X$(Na) & Na/H & $5.2\times10^{-6}$ \\
\hline 
 Y$_0$(H$_2$O) & H$_2$O/H & 0.9 $X$(O) \\
 Y$_0$(CO) & CO/H & 0.1 $X$(O) \\
 Y$_0$(CH$_4$) & CH$_4$/H & 0.84 $X$(C) \\
 Y$_0$(C) & C/H & $8.5\times10^{-3}$ $X$(C) \\
\hline
\end{tabular}
\end{table}
\null

\begin{figure}
\centering
\includegraphics[scale=0.4]{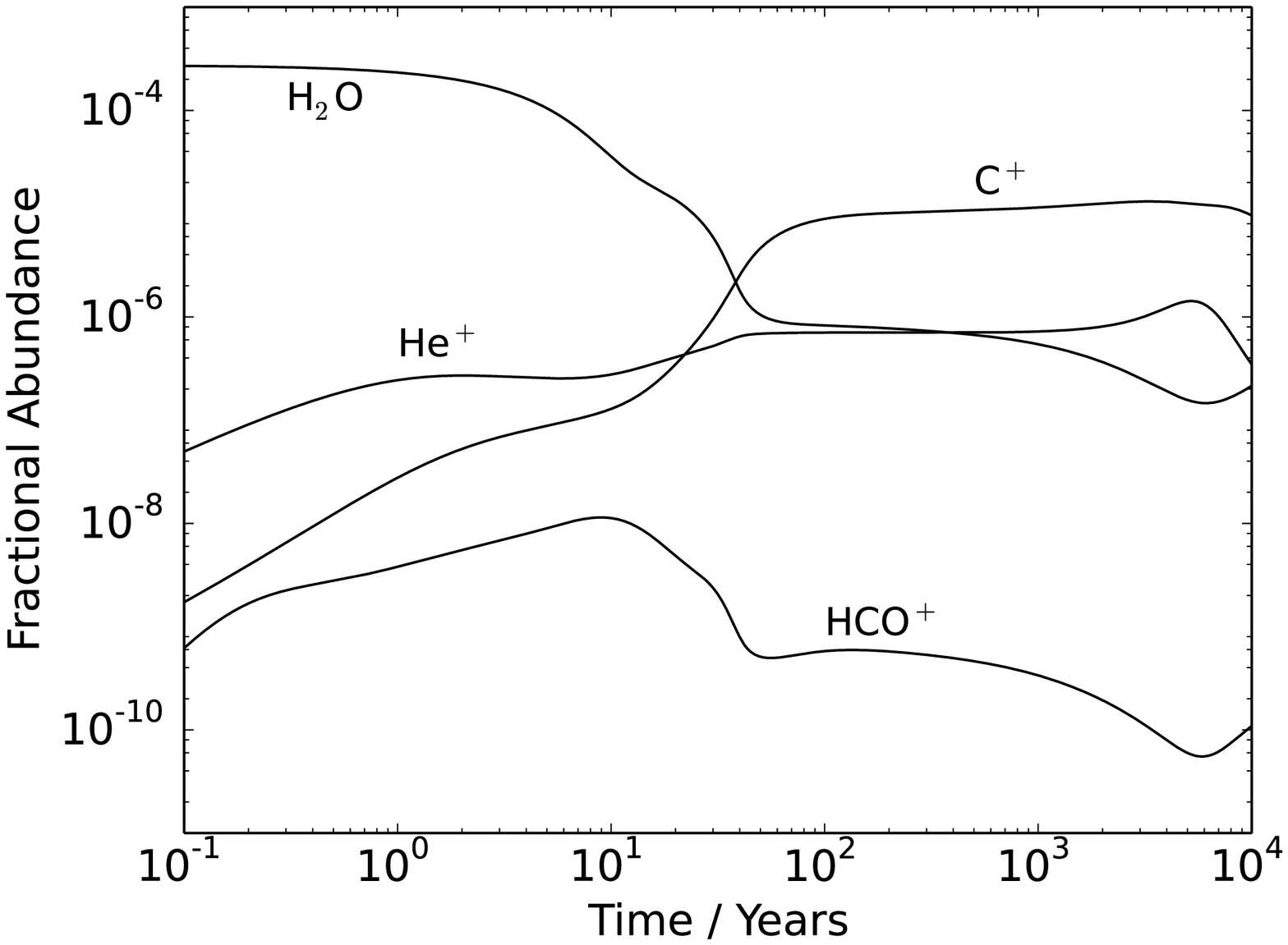}
\includegraphics[scale=0.4]{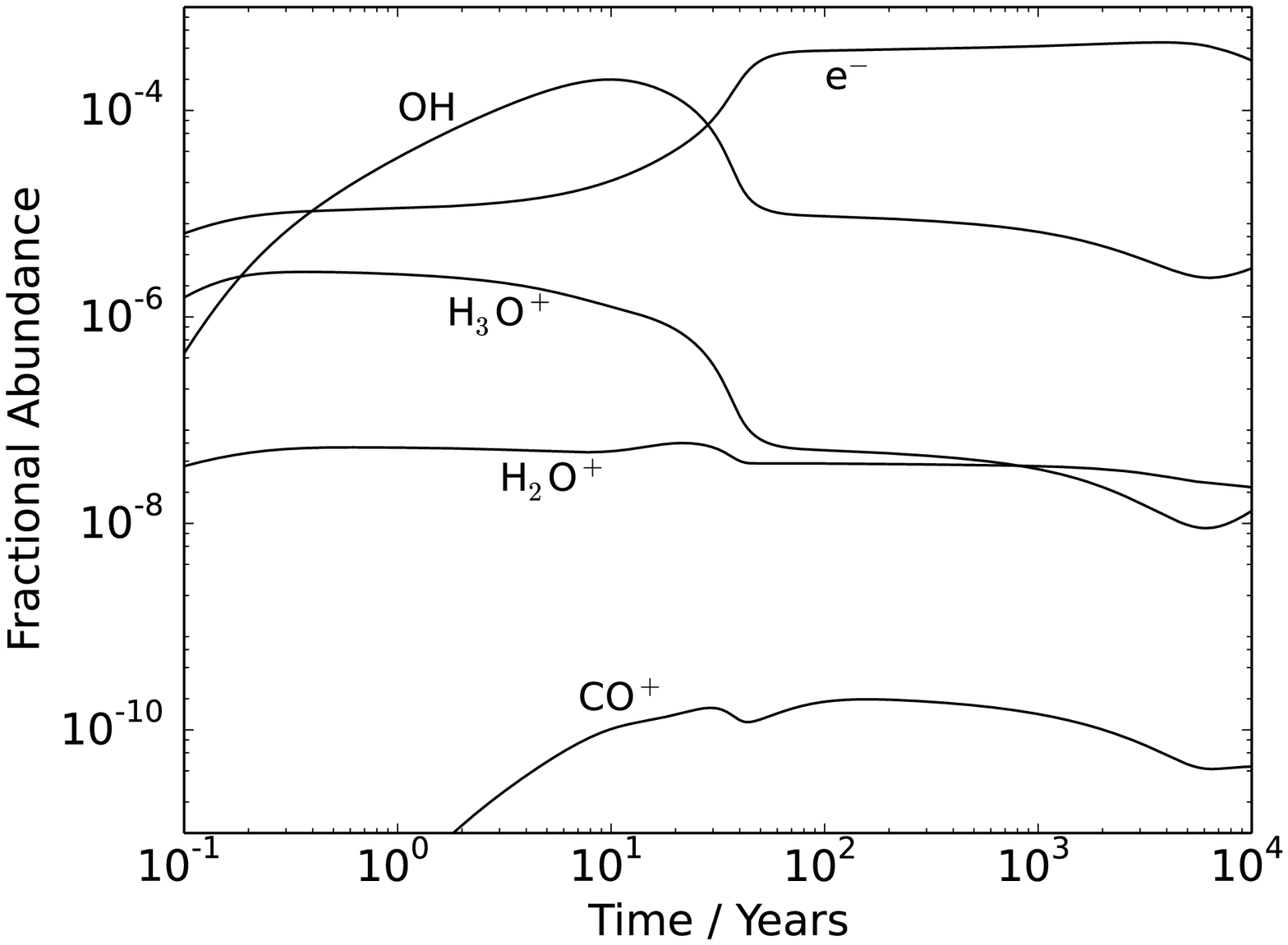}
\includegraphics[scale=0.4]{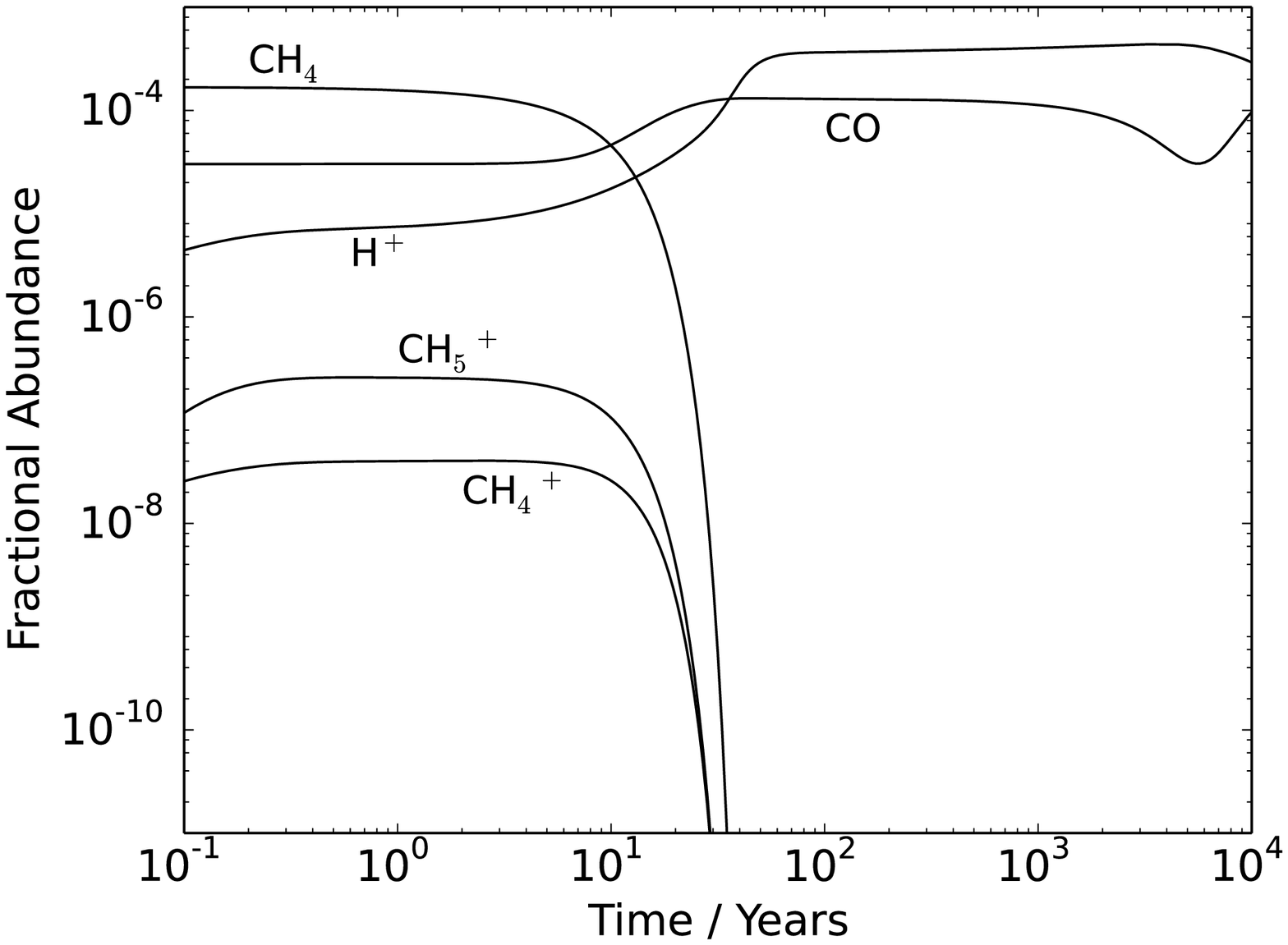}
\caption{Modelled fractional chemical abundances relative to hydrogen near the base of a cavity of age $11~{\rm kyrs}$ as a function of chemical integration time. The interface temperature is $150~{\rm K}$ and the density is $10^5~{\rm cm^{-3}}$. The abundance of HCO$^+$ is seen to correlate with a variety of parent molecules including H$_2$O, OH and CH$_4$ before they are all destroyed. A quasi--equilibrium is reached after approximately $1~{\rm kyr}$.}
\label{fig:formation}
\end{figure}

The abundances of the chemical species are allowed to evolve with the gas phase reactions given in the RATE06 database \citep{Wea:2007}. In addition, H$^+$, C$^+$ and O$^+$ are continuously injected at a rate given by Equation \ref{eq:injection} multiplied by the appropriate elemental abundance and using the parameters in Table \ref{table:outflowparams}. Helium is also injected and, following previous studies \citep[e.g.][]{RTW:2000}, is assumed to be only $10\%$ ionized. Sufficient electrons to maintain overall charge neutrality are also added to the interface. The cosmic ray ionization rate is set to a standard value of $\zeta = 10^{-17}~{\rm s^{-1}}$. We set the UV radiation field in our model interface to zero throughout this paper, considering only cosmic ray induced UV photoprocesses. This allows us to compare results of our hot ion chemistry model with those of the photochemistry model \citep{RTW:2000}. \cite{RTW:2000} argue that UV photons are created in the bow shocks of the outflow. In fact, the leading initial bow shock of a jet should have propagated a long way from the regions near the central source we are examining. Instead, such jets usually have internal working surfaces that are weaker shocks caused by variations in the jet mass loss rate and speed. These weaker shocks and working surfaces should form a reasonably constant source of high temperature and local UV radiation that can ionize the outflow material near the star but with negligible effect on the interface. \cite{vKea:2009} argue that for the source HH 46 the radiation field reaching the cavity wall could be as high as 600 Draine. However, factors including weaker shocks in different sources, photoionization of outflowing gas within the cavity and geometric variations along the cavity wall could all significantly reduce this, especially close to the central source. We use this argument to justify our assumption that the outflow material is ionized while there is negligible UV radiation in the interface itself.

The density and temperature of the interface are poorly understood and are treated as free parameters. A number of works have attempted to infer these properties using observations, models and simulations and can guide us in our choice. \cite{Cea:2006} simulate turbulent interfaces with densities of $10^3$--$10^5~{\rm cm^{-3}}$, while the magnetohydrodynamic simulations of \cite{Bea:2011} and \cite{BTP:2014} are capable of much higher densities (although at earlier times). Temperatures of order $100~{\rm K}$ are also seen in the last of these works. Observational studies such as \cite{vKea:2009} suggest relatively low temperatures in the range of 80--150 K and densities of a few times $10^4~{\rm cm^{-3}}$ for the HH46 outflow, although there are uncertainties since the observed CO 3--2/6--5 line ratio is not expected to change significantly as the kinetic temperature increases above $100~{\rm K}$ \citep{vdTea:2007}. These contrast with the predictions of entrainment models where outflow temperatures of over $1000~{\rm K}$ are possible (\citealt{Cea:2006}, \citealt{LWM:2013}).

In Section \ref{sec:results} we choose to consider a grid of densities from $3\times10^4$ to $10^6~{\rm cm^{-3}}$ and temperatures from $100$ to $150~{\rm K}$ for all reactions. These ranges represent a suitable intermediate between the cold, dense molecular envelope and the hotter, more diffuse outflowing material as they mix turbulently with one another. While high--J CO \citep{Yea:2010} and HCO$^+$ \citep{Kea:2013} observations would seem to demand a much hotter interface, it is possible that such transitions could be collisionally excited by energetic electrons from the injecta due to the typically lower critical densities. In fact, our typical equilibrium ionization fractions of a few times $10^{-4}$ are close to the prediction by \cite{JSea:2006} of $10^{-3}$ in the magnetic precursor of an outflow C--shock from measurements of the excitation of H$^{13}$CO$^+$, among other species. We also note that if temperatures were significantly less than $100~{\rm K}$, freeze out of water onto grains would remove the main reagent for the formation of HCO$^+$ and a non--thermal desorption mechanism would be required to liberate the water ice. While other models may appeal to photodesorption by direct UV radiation, it may also be possible to initiate our model with secondary UV photons due to ionization by cosmic rays in such a case.

Figure \ref{fig:formation} shows chemical abundance profiles as a function of time for a typical parameter set. As with the photochemical pathway presented in \cite{RTW:2000}, our injection model can give an enhancement in the HCO$^+$ abundance to $\approx10^{-8}$, but only on timescales of 10--100 years. Beyond that time, dissociative recombination becomes significant and a lower abundance remains. An approximate chemical equilibrium is reached by the system on timescales of the order 100 years for all species, roughly independent of the physical parameters. With reference to Figure \ref{fig:timescale}, for a cavity of age over $10~{\rm kyrs}$ such as B5 IRS1, the injection evolution timescale is more than $1~{\rm kyr}$ on all angular scales and so the interface can reach this equilibrium. We are therefore justified in using a chemical integration time of $1~{\rm kyr}$ to represent the equilibrium state of older cavities like B5 IRS1. For younger cavities, this injection evolution timescale becomes shorter and so a shorter chemical integration time will be necessary. Note that small abundance variations are seen at approximately $10~{\rm kyrs}$ due to reactions with He$^+$, but these are only temporary fluctuations lasting a few ${\rm kyrs}$.

As in the work of \cite{RTW:2000}, the main pathway to HCO$^+$ formation on timescales of approximately 10 years is the reaction of water from the initial chemical composition of the gas with ionised carbon. The material injected into the interface is a much weaker source of ionised carbon than photoionisation, meaning this reaction alone does not produce as much HCO$^+$ in our model. However, two other routes to HCO$^+$ are initiated by hydrogen ions present in the injected material in our model but typically absent in photochemical reaction networks. The first involves the formation of the CH$_5^+$ ion which can then react with carbon monoxide:

\[ {\rm CH_4 + H^+ \rightarrow CH_4^+ + H}\] 
\[ {\rm CH_4^+ + H_2 \rightarrow CH_5^+ + H}\] 
\[ {\rm CH_5^+ + CO \rightarrow HCO^+ + CH_4.}\] 

\noindent The other pathway involves the formation of CO$^+$ ions, either by the cosmic ray ionization of carbon monoxide or the reaction of OH with injected carbon ions:

\[ {\rm H_2O + H^+ \rightarrow H_2O^+ + H}\] 
\[ {\rm H_2O^+ + H_2 \rightarrow H_3O^+ + H}\] 
\[ {\rm H_3O^+ + e^- \rightarrow OH + H + H}\] 
\[ {\rm OH + C^+ \rightarrow CO^+ + H.}\] 

\noindent CO$^+$ then readily reacts with available molecular hydrogen to produce HCO$^+$:

\[ {\rm CO^+ + H_2 \rightarrow HCO^+ + H.}\] 

\noindent These three reaction pathways together lead to abundances of HCO$^+$ that are comparable to the results of \cite{RTW:2000}; typically between 10$^{-7}$ and 10$^{-9}$ relative to hydrogen in the first 100 years of chemical integration.

\section{Results}
\label{sec:results}

\begin{figure*}
\centering
\includegraphics[scale=0.8]{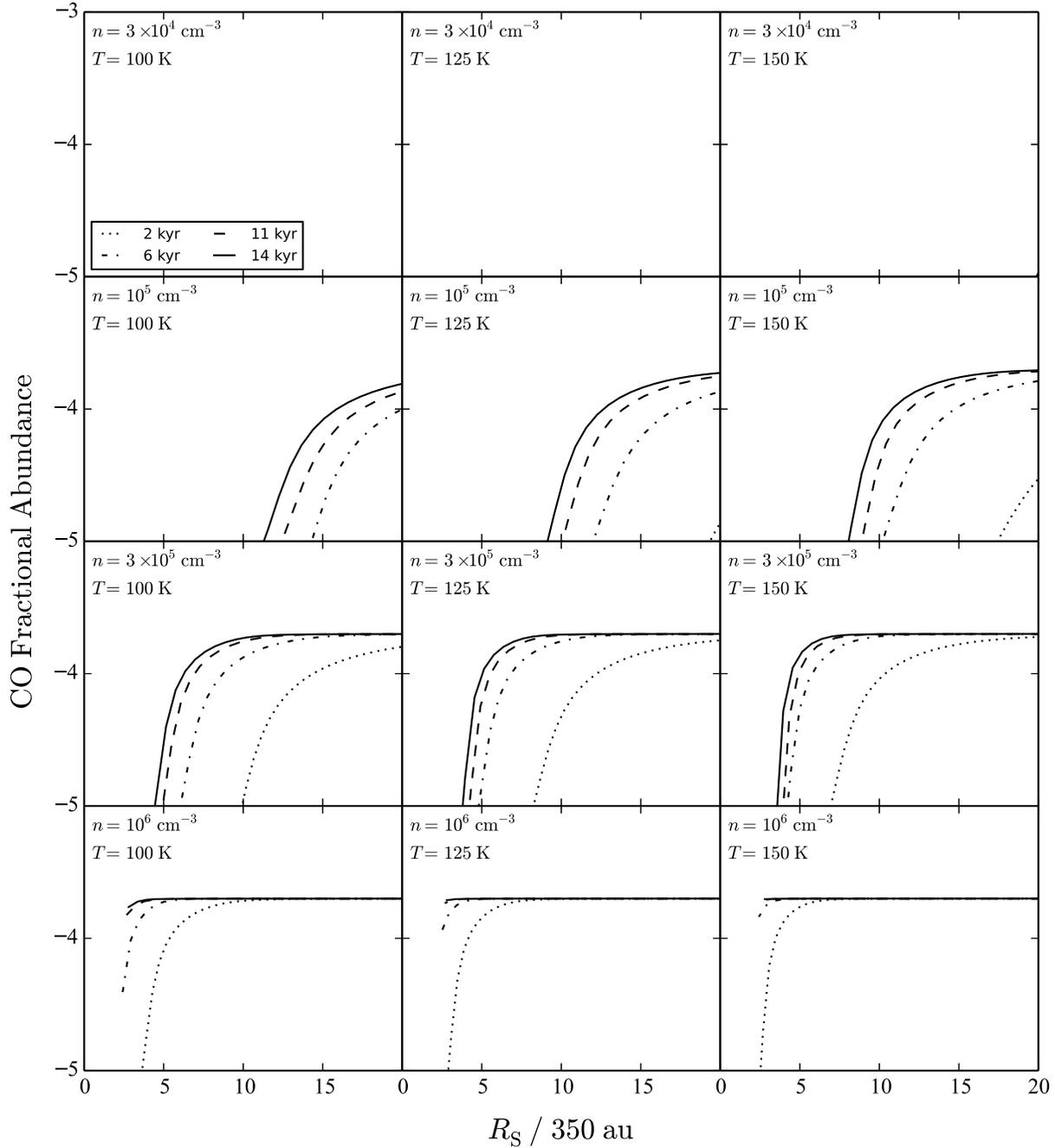}
\caption{Fractional abundances of CO relative to hydrogen for a cavity of dynamical age 2 (dot), 6 (dot--dash), 11 (dash) and 14 (solid)$~{\rm kyrs}$ as a function of distance from the central source. The temperatures used are $100~{\rm K}$ (left), $125~{\rm K}$ (centre) and $150~{\rm K}$ (right). Densities are fixed at $3\times10^4~{\rm cm^{-3}}$ (top) through $10^5~{\rm cm^{-3}}$, $3\times10^5~{\rm cm^{-3}}$ and $10^6~{\rm cm^{-3}}$ (bottom). The chemical integration time at which the abundances are plotted are 100, 300, 1000 and 1000 years for the 2, 6, 11 and 14 $~{\rm kyrs}$ cavities respectively.}
\label{fig:multiCO}
\end{figure*}

\begin{figure*}
\centering
\includegraphics[scale=0.8]{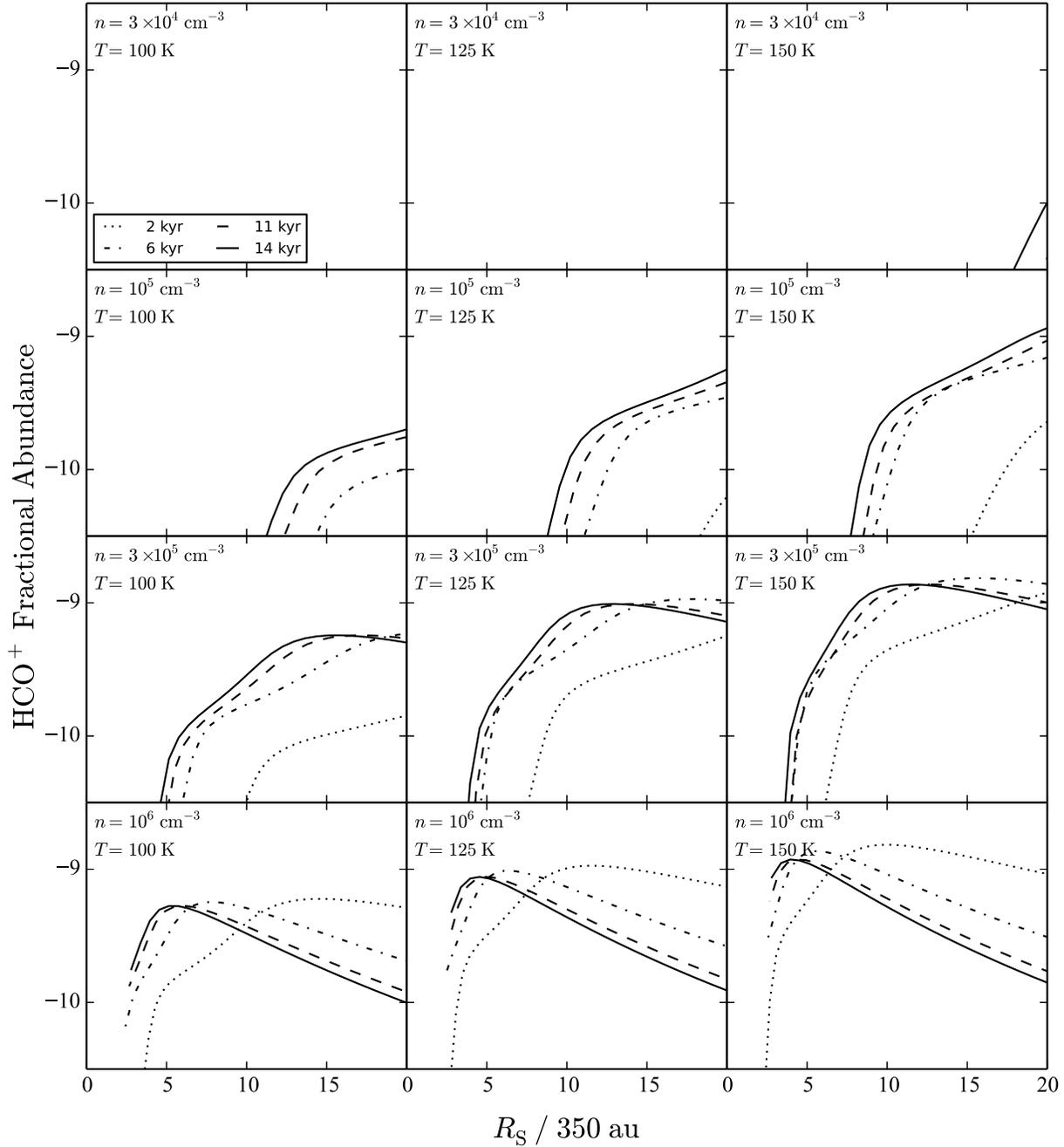}
\caption{The same as Figure \ref{fig:multiCO} except showing the fractional abundances of HCO$^+$ relative to hydrogen.}
\label{fig:multiHCO}
\end{figure*}

\begin{figure}
\centering
\includegraphics[scale=0.66]{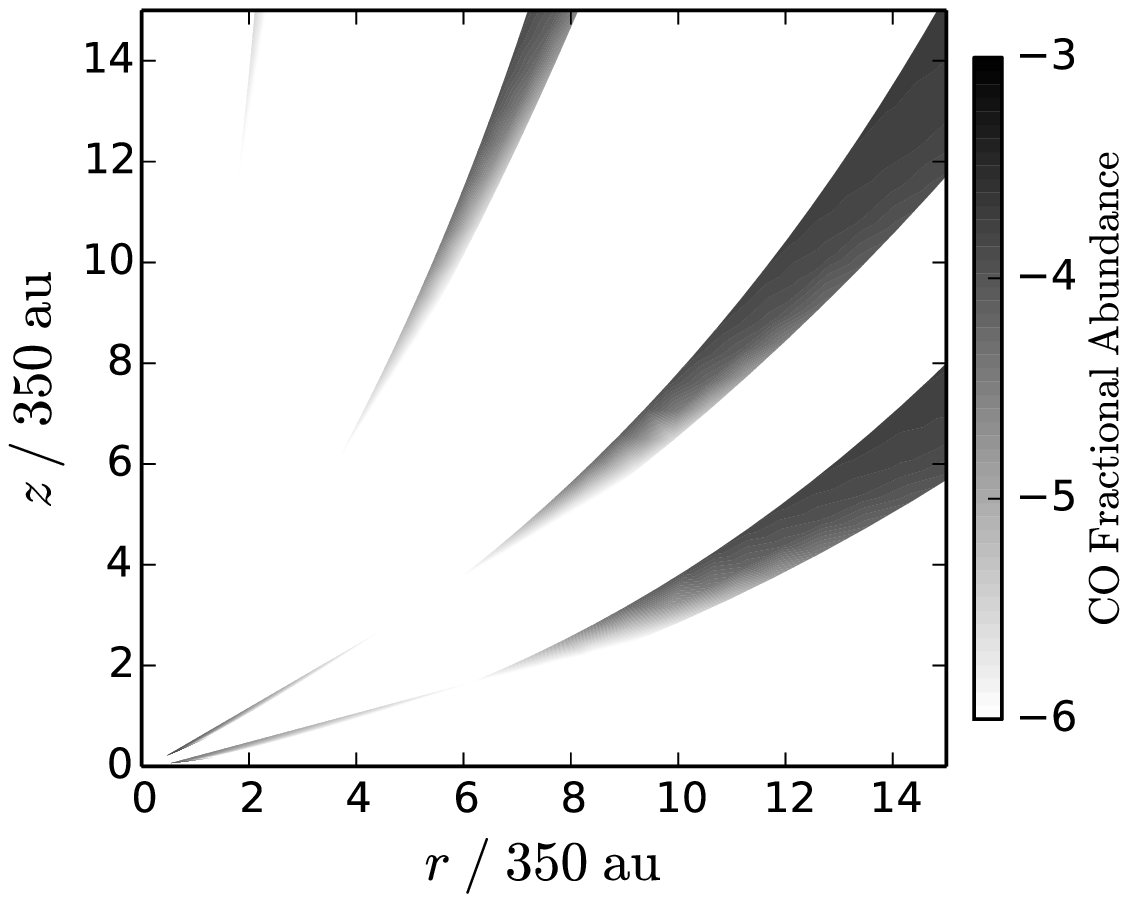}
\includegraphics[scale=0.66]{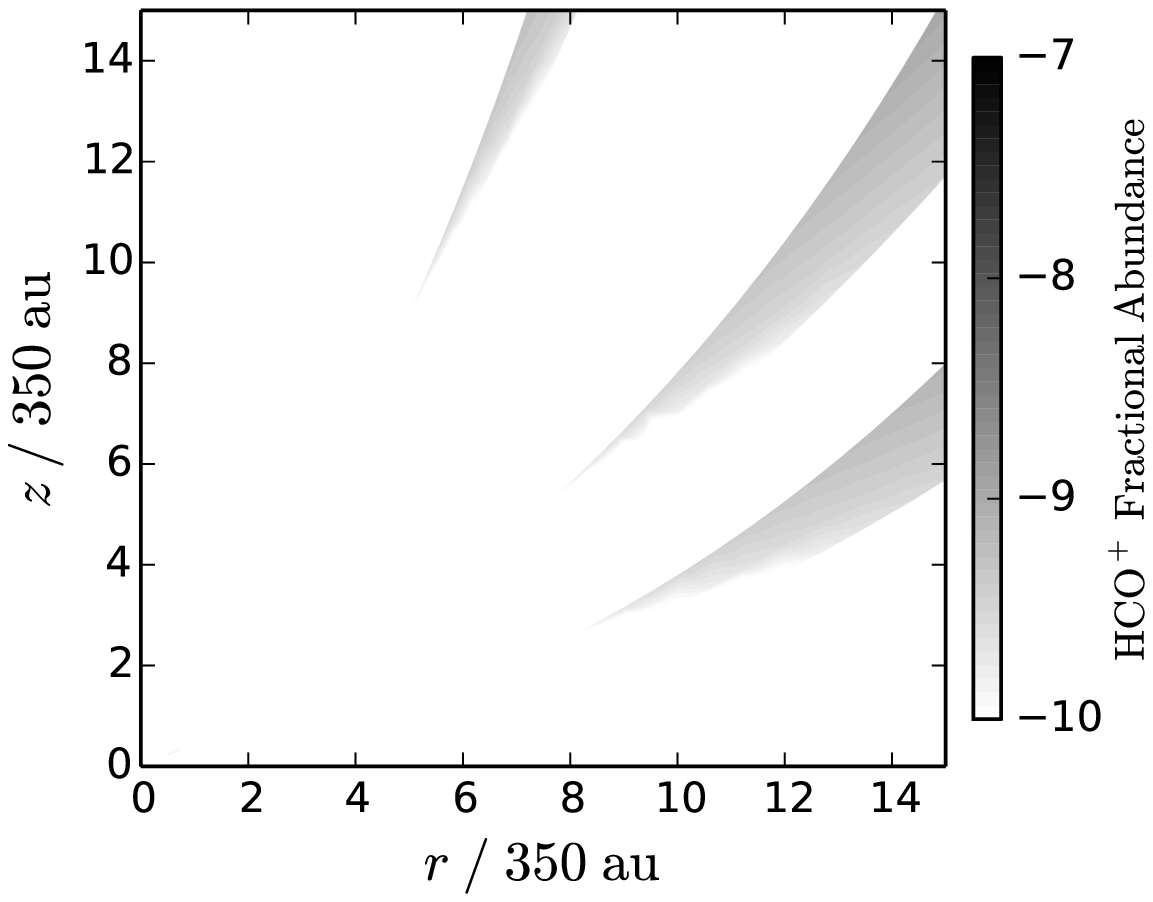}
\caption{Cross--sectional fractional chemical abundance maps of CO (top) and HCO$^+$ (bottom) for cavities of dynamical age 2 (narrowest), 6, 11 and 14 (widest) kyrs. The interface density is $3\times10^5~{\rm cm^{-3}}$ and the temperature is $T=150~{\rm K}$ giving the best qualitative fit to the observed chemical properties. The chemical integration time at which the abundances are plotted are 100, 300, 1000 and 1000 years for the 2, 6, 11 and 14 $~{\rm kyrs}$ cavities respectively. The interface thickness has been exaggerated to 50\% of the dynamical radius for clarity. Note that the youngest cavity is not seen in either plot due to the fractional abundances of both species being off the bottom of their respective scales.}
\label{fig:maps}
\end{figure}

In this section we present the grid of outputs from our model for a range of parameters. First, we took four cavity dynamical ages ($t_{\rm dyn}$ = 2, 6, 11, $14~{\rm kyrs}$) and calculated the turbulent injection rate (Equation \ref{eq:injection}) as a function of position. We then integrated the resulting chemistry for a chemical age, $t_{\rm chem}$, \text{of} up to $10~{\rm kyrs}$ at all positions and using a grid of densities ($n = 3\times10^4$ to $10^6~{\rm cm^{-3}}$) and temperatures ($100~{\rm K}$, $125~{\rm K}$ and $150~{\rm K}$). The spatial abundance profiles of the interfaces in CO and HCO$^+$ are presented in Figures \ref{fig:multiCO} and \ref{fig:multiHCO} respectively. The $2~{\rm kyrs}$ interface is plotted after a chemical integration time of 100 years, the $6~{\rm kyrs}$ interface after 300 year and the older 11 and $14~{\rm kyrs}$ interfaces after a chemical age of $1~{\rm kyr}$; motivated by the results of Figure \ref{fig:timescale} and Section \ref{sec:chemistry}.

A number of features are apparent and allow us to constrain the physical conditions of the interface. Over the observed range of parameters, the chemical composition is more strongly dependent on density than on temperature. Below densities of $10^5~{\rm cm^{-3}}$, carbon monoxide is unable to saturate due to  the dissociation by injected He$^+$:

\[ {\rm He^+ + CO \rightarrow C^+ + O + He.}\] 

\noindent At $10^5~{\rm cm^{-3}}$ a transition to CO saturation is seen, with the regions of the interface at larger radii (and hence lower ion injection rates) able to sustain a significant amount of carbon monoxide. With higher densities, the injected material has a lower impact on molecular abundances and saturation is achieved for all but the regions closest to the central source. The CO dependence on temperature is minimal over this range. What is of most interest is that at the density of $10^5~{\rm cm^{-3}}$ there is a marked variation in CO abundance patterns over the age of the cavity. The youngest, narrowest interface modelled has a severely diminished carbon monoxide fractional abundance of less than 10$^{-6}$ over all angles. By comparison, the $6~{\rm kyrs}$ cavity exhibits a small band of CO saturation at radii greater than approximately 4000~{\rm au}. The two older and wider angle interfaces show a significantly increased band of CO saturation in to closer radii. We argue in Section \ref{sec:discussion} that this variation is not sensitive to the choice of chemical timescales at each cavity age. 

This feature of more CO at wider angles in the model can be compared with the observational preference towards wide angled outflows observed in CO. It raises the possibility that if the dynamics of protostellar outflows causes a decollimation effect then there could be an observational bias against the youngest of these objects due to the lower abundance of carbon monoxide. This would reproduce the apparent observational selection of wide angled ``brandy snifter'' shaped outflows in preference to narrower ``champagne flute'' shapes as noted in Section \ref{sec:intro}. However, at the smallest radii, the model yields negligible CO abundances at all ages, failing to reproduce the observations that continue all the way in to the central object. Since our injection function (Equation \ref{eq:injection}) scales as $R_{\rm S}^{-3}$, the density of hot ions is greatest nearest the central source and as such, helium ions more readily dissociate any carbon monoxide present in this region. Potential resolutions to this issue are considered in Section \ref{sec:discussion}.

The modelled HCO$^+$ abundance distributions exhibit similar properties to CO. Below the density of $10^5~{\rm cm^{-3}}$ almost none is able to form, while fractional abundances of up to $\approx$ 3$\times$10$^{-9}$ can form above this threshold. HCO$^+$ abundances are also marginally higher at higher temperatures due to the reduced dissociative recombination rate. Furthermore, HCO$^+$ is more prominent at larger radii due to the reduced electron injection rate leading to a slower rate of dissociative recombination.
At densities of $3\times10^5~{\rm cm^{-3}}$ and $10^6~{\rm cm^{-3}}$ there is also evidence for a preferential distance (or equivalently polar angle) at which the abundance of HCO$^+$ peaks in equilibrium. Whether or not such a feature would be observable is unclear, but it does show the fine balance between needing enough H$^+$ from the injecta to form a large fractional abundance of HCO$^+$ in equilibrium and not adding too many electrons to destroy it all.

The features of the model discussed above suggest that the physical conditions leading to the qualitative observations of abundant carbon monoxide and HCO$^+$ are quite finely tuned. The combination of density $n=3\times10^5~{\rm cm^{-3}}$ and temperature $T=150~{\rm K}$ appears to give the best qualitative match. Such parameters are in broad agreement with the various observations, models and simulations discussed in Section \ref{sec:chemistry}. The abundance cross-sectional maps of CO and HCO$^+$ for that combination of parameters is shown in Figure \ref{fig:maps}. Most strikingly, the youngest 2 kyr cavity fails to reach CO saturation and would therefore likely be unobservable, in agreement with the hypothesis that the morphological bias towards wide--angled outflows is chemical in origin. While the equilibrium abundances of HCO$^+$ shown in the figure are somewhat lower than observed, it is still true that a short lived burst in fractional abundance is seen on timescales of approximately 10 years as shown previously in Figure \ref{fig:formation} and the same as in the previous photochemical model \citep{RTW:2000}.

\section{Discussion}
\label{sec:discussion}

\begin{figure*}
\centering
\includegraphics[scale=0.8]{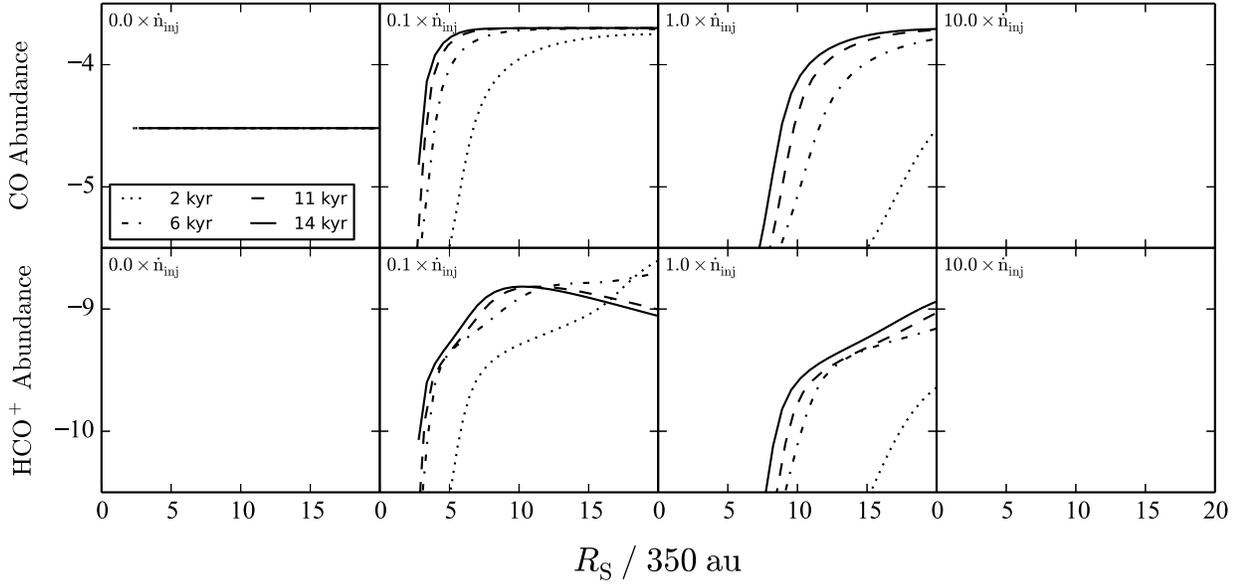}
\caption{Fractional abundances of CO (top) and HCO$^+$ (bottom) relative to hydrogen for a cavity of dynamical age 2 (dot), 6 (dot--dash), 11 (dash) \& 14 (solid)$~{\rm kyrs}$. The temperature is fixed at $150~ {\rm K}$ and the density at $10^5~{\rm cm^{-3}}$. A multiplicative prefactor to the injection rate of Equation \ref{eq:injection} is used, ranging from 0 (left) through 0.1, 1 and 10 (right). The chemical integration time at which the abundances are plotted are 100, 300, 1000 and 1000 years for the 2, 6, 11 and 14 ${\rm kyrs}$ cavities respectively.}
\label{fig:inj}
\end{figure*}

\begin{figure*}
\centering
\includegraphics[scale=0.8]{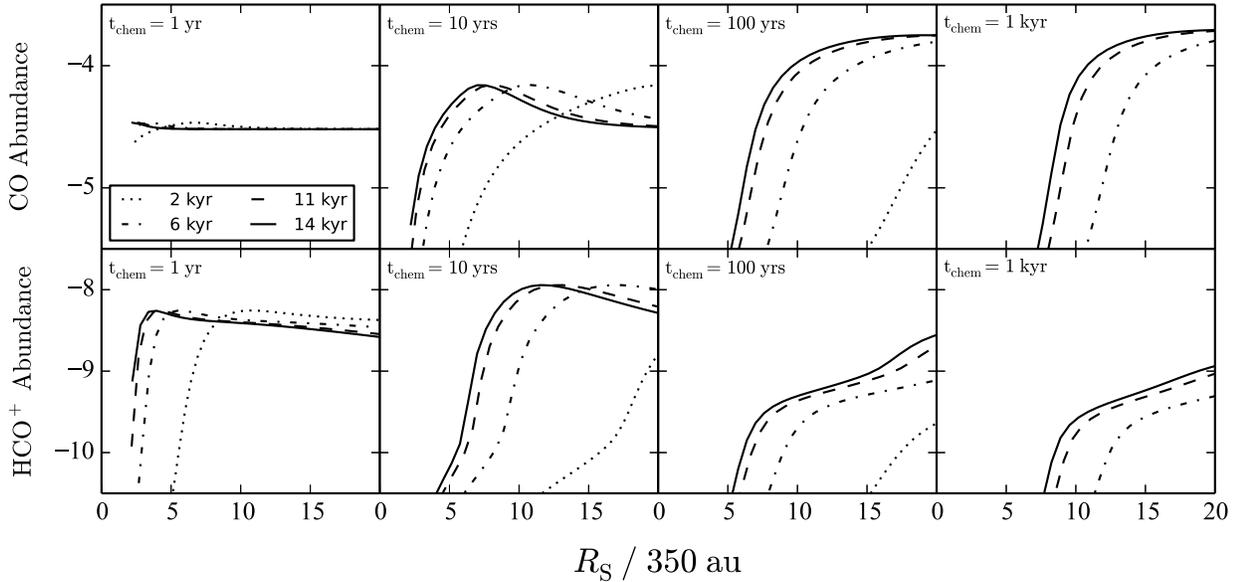}
\caption{Fractional abundances of CO (top) and HCO$^+$ (bottom) relative to hydrogen for a cavity of age 2 (dot), 6 (dot--dash), 11 (dash) \& 14 (solid)$~{\rm kyrs}$. The temperature is fixed at $150~{\rm K}$ and the density at $10^5~{\rm cm^{-3}}$. The chemical integration time is varied from 1 year (left) through 10, 100 and 1000 years (right).}
\label{fig:chemAge}
\end{figure*}

We have produced a model for the chemistry in protostellar outflow interfaces driven by injected ions that is able to broadly reproduce some of the quantitative and qualitative features of those systems. For the right choice of density, bands of carbon monoxide are observed in the edges of cavities with opening angles larger than $\approx50^\circ$. For a cavity with a symmetry axis inclined in the plane of the sky, this would yield column densities of the order a few times $10^{18}~{\rm cm^{-2}}$ which is easily observable. Furthermore, the critical density for CO(2--1) observations as taken from the Leiden Atomic and Molecular Database \citep{Sea:2005} is approximately $10^5~{\rm cm^{-3}}$. This value of critical density supports the view that interfaces with this density should be readily observable and, according to the model presented here, should show a wide--angled morphology. HCO$^+$ is also produced in our model, although not in quite the quantities inferred from observations. However, temporary enhancements to 10$^{-8}$ occur on timescales of 10--100 years as in \cite{RTW:2000}.

One key shortcoming is the inability to form carbon monoxide all the way into the central source. The cubic radial dependence $\propto$$R_{\rm S}^{-3}$ was the main factor in determining the local ion injection rate and not $\sin\psi$ as predicted. Due to this strong dependence on radius, CO is rapidly destroyed close to the opening angle due to too much He$^+$ being injected. There are, however, a number of potential refinements that may correct this deficiency. The first point of note is that while the ejected material is presumed to be uniform with solid angle in our model, it could be argued that the presence of any magnetic field (for example) may collimate it, with more material ejected towards the poles, so that the injection rate at wide angles is reduced. This could take the form of a further geometric factor in the injection rate equation, e.g. $\propto\cos(\theta/\theta_{\rm m})$; maximal near the pole, zero at the opening angle and normalised over solid angle. Another consideration is that the cavity wall thickness may not be proportional to the dynamical radius. In their simulations, \cite{Cea:2006} observe interfaces that are not only approximately constant thickness as a function of angle, but in fact grow with time. While the way in which this growth may be parametrised is not clear, it is evident that the thickness of the interface in our model could have been underestimated close to the central source. Increasing it would have the effect of diluting the injected helium ion abundance and hence reducing the rate of CO destruction. Finally, while reasonably argued from geometric considerations, the turbulent mixing efficiency $\sin\psi$ is surely uncertain. If it were a stronger function of the mixing angle (for example, $\sin^2\psi$), this would have the effect of again reducing the injection rate near the central source where the mixing angle $\psi\rightarrow0$ and could yield an enhanced CO abundance.

The model features mentioned above are not considered explicitly in this exploratory paper, rather reserved for future work. However, variations in some of the parameters that were considered fixed in previous chapters can shed some light on the response of the model that can be expected. The simplest of these is to allow a global multiplicative pre--factor to scale the injection rate in Equation \ref{eq:injection}. The results of varying this pre--factor are summarised in Figure \ref{fig:inj}. Reassuringly, on the timescales we considered before, our initial conditions are in equilibrium in the absence of injection, while scaling the injection rate up by a factor of 10 removes both CO and HCO$^+$ from the system on all angular scales of interest. However, reducing the injection rate (as given by Equation \ref{eq:injection}) by a factor of ten demonstrates quite how
sensitive
the system is to the rate of turbulent injection. The interface then contains somewhat more HCO$^+$ on all angular scales and CO is observed in to much closer radii, although the discrimination between CO abundances in older and younger cavities is lost. All of the issues discussed above -- interface thickness, mixing efficiency and collimation of the ejecta -- could be realised in terms of a reduction in the injection efficiency, particularly close to the central source. 

One other choice of parameter that was discussed thoroughly in Sections \ref{sec:cavity} and \ref{sec:chemistry} is the chemical integration time, $t_{\rm chem}$. This choice was made to be shorter than the typical timescale on which the injection rate changes for its age. However, as shown in Figure \ref{fig:timescale} the timescales near the cavity base are significantly shorter than further out; this was not considered in Section \ref{sec:results}. Figure \ref{fig:chemAge} shows the evolution of the interface carbon monoxide and HCO$^+$ fractional abundances with chemical integration time. The relative stability of CO and the flare up of HCO$^+$ between 1 and 10 years of chemical integration is evident. However, between 100--1000 years, the profiles of both remain roughly constant, with only a small amount lost to the usual reaction channels. Figure \ref{fig:timescale} shows that the chemical age is greater than 100 years for all angles and cavity ages. Therefore, we argue that the choice of chemical age between 100--1000 years from Section \ref{sec:results} has limited effect on the overall results. 
From Figure \ref{fig:formation} we can see that the abundances of both 
CO and HCO$^+$ dip at t$\sim$6,000 years. This {\em could} inhibit emission
from some older outflow cavities, suggesting that, rather than just a minimum,
there may be a range of cavity opening angles that are conducive to observable
emission. However, the abundances recover at later times, so the effect may be
marginal.

As a part of their work to model the abundance of CO$^+$ in outflow systems, \cite{Bea:2009} considered a similar mixing mechanism driving the chemistry. While their model injected only ionised atomic hydrogen and electrons, they concluded it could only produce column densities $N(CO^+)\approx10^8~{\rm cm^{-2}}$; some two orders of magnitude less than their two dimensional photochemical model that matches the observational data. While our chemical scheme produces comparable peak abundances ($Y(CO^+)\approx10^{-10}$), in some regions this abundance is sustained for as long as $10^{4}~{\rm yrs}$ by comparison with less than one year in their model. We can therefore assume in our model that CO$^+$ is present through the full width of our cavity wall giving column densities for a cavity in the plane of the sky as high as $N(CO^+)\approx10^{12}~{\rm cm^{-2}}$. Although this maximal value may be larger than the inferred column density, it is possible that our full injection model can provide an adequate explanation of the observed CO$^+$ signal. The high peak abundance ($Y(CO^+)>10^{-4}$) as well as the strong correlation with HCO$^+$  over the chemical integration time make it an interesting species for comparison with other models. Another chemical signature of our model is the high abundance of OH seen in Figure \ref{fig:formation}. While previous studies have not provided a prediction on the presence of the OH radical, it would be expected in photochemical models following the photodissociation of water.

A number of potential chemical mechanisms were neglected in our modelling that could have an effect on the numerical results of our study. Importantly, the presence of small dust grains and polycyclic aromatic hydrocarbons (PAHs) is believed to reduce the abundance of free electrons by combining with them rapidly to form negatively charged particles. This could significantly reduce the rate for dissociative recombination of HCO$^+$ by electrons; the dominant destruction pathway. While the negatively charged PAHs can themselves react to dissociate HCO$^+$, they move significantly slower than electrons and so the rate coefficient for that process will be significantly lower than with electrons, leading to an expected increase in the HCO$^+$ abundance in the presence of PAHs. This argument would suggest that the HCO$^+$ abundances in our model are therefore underestimates of the true column densities. On the other hand, negatively charged PAHs provide further chemical pathways for recombination, such as mutual neutralization reactions with C$^+$ ions \citep{BT:1998}. This could reduce the formation rate of HCO$^+$ meaning that our modelled fractional abundances are actually overestimates. Since the model of \cite{Bea:2009} also contained PAH chemistry, such reactions might also provide an explanation for the significant drop in CO$^+$ abundances that they found after one year. In the case of high mass young stellar objects, \cite{SDvDB:2005} demonstrated that the presence of X--ray photons from the central source acted to provide an enhancement in the HCO$^+$ abundance due to more C$^+$ ions produced by secondary photons from excited molecular hydrogen. Low mass analogues are expected to have similar X--ray luminosities \citep{M:2001} and so we expect that the presence of X--rays would complement our model by enhancing the HCO$^+$ abundance while leaving CO abundances roughly the same.

We also note that the origin of the cavity is not explicitly considered. Whilst we adopt the model of a decollimating outflow of constant total mass loss rate as in \cite{CRW:2008}, the physics leading to such a system was never considered; the model was simply fit to the observed shape to constrain the physical parameters of the ejecta and envelope. In fact, the model of \cite{LWM:2013} is static by comparison, with the ram pressure of the outflow balanced by the turbulence in the interface. It is possible to apply our model to such static systems by considering the full chemical equilibrium as $t_{\rm chem}\rightarrow\infty$. Once again, Figure \ref{fig:chemAge} suggests that this would have the effect of reducing the HCO$^+$ abundance but also removing most CO from the two younger cavities. This then fits even better with the observed bias, with only more powerful outflows that lead to wider opening angles being observable. Furthermore, our ignorance of the dynamics of the mixing in the cavity wall means we are unable to comment on the observed broad line profiles of CO \citep{Yea:2012} and HCO$^+$ \citep{Kea:2013} at high--J. While a shock origin has been proposed, a more detailed dynamical model of the mixing in the cavity walls may tell us if a turbulent origin may also be valid.

Finally, it is noted that a two dimensional photochemical model could produce a similar morphological bias. However, the main source of UV radiation in the cavity is uncertain. In their model, \cite{Vea:2012} considered the central protostellar object as a source of isotropic UV radiation, meaning the local radiation field varies along the cavity wall in a similar manner to our ion injection rate due to the same geometric spreading and impact angle. The resulting chemistry is therefore likely to have a qualitatively similar spatial dependence to our turbulent mixing model. The same is also likely to be true for UV photons originating from the accretion luminosity of the inner protostellar accretion disk \citep{Sea:1995}. However, \cite{Vea:2012} also suggest local shocks in the cavity wall as it expands into the surrounding cloud could be a source of UV radiation that is more uniform along the cavity wall leading to a more uniform abundance profile. Furthermore, shocks from the the internal working surfaces of the optical jet (\citealt{RTW:2000}, \citealt{vKea:2009}) could lead to the local radiation field and hence chemical abundances in the cavity wall being enhanced in the vicinity of knots in the jet. The spatial variations in the CO and HCO$^+$ abundances should be uniquely sensitive to the morphology of the system driving the chemistry, whether driven by photochemistry or turbulent mixing of ions.

\section{Conclusions}
\label{sec:conclusions}

We have developed a model for the chemistry at the interface between the outflow and the molecular envelope for low mass protostars. The chemistry is driven by turbulent mixing of hot ions from the ejecta with cold molecular gas in the envelope. In particular we observe the following features of the model:

\begin{itemize}

    \item A preference towards systems with wider opening angles in the modelled CO abundances.

    \item High abundances of HCO$^+$ without the need for photochemistry.

    \item Approximate chemical equilibrium reached on the timescale at which the injection rate changes, leaving the results independent of the choice of chemical integration times greater than 100 years.

    \item The results are acutely sensitive to the interface density, with $10^5~{\rm cm^{-3}}$ providing the best  fit to observed CO morphologies.

    \item Temperatures of $150~{\rm K}$ give the maximum possible stable HCO$^+$ abundance within the range investigated.

    \item The assumed magnitude of the injection rate is critical in determining if CO and HCO$^+$ form or are destroyed by injected He$^+$ and e$^-$ respectively.

\end{itemize}

These results are quantitatively similar to those of \cite{RTW:2000} using photochemistry, with the addition that observed CO maps are qualitatively matched by the model abundances. This provides a possible explanation for the observational preference towards wide opening angle outflows:  higher injection rates of dissociating ions for narrower opening angles reduce the CO abundance and make it unobservable. A three dimensional radiative transfer treatment would allow us to consider the degree to which a realistic abundance profile affects the inferred CO map, and to determine if such a treatment is necessary to understand the observations. It would also be interesting to consider photochemistry and ion driven chemistry in these systems side by side and in combination, to look for unique tracers that might reveal the true nature of the chemistry at work. We reserve these considerations for future work.

\section*{Acknowledgements}

The authors wish to thank the anonymous referee for their in--depth comments that let to significant clarification of the text and main results, as well as pointing out a number of previous works in the field that provided useful insight and direct comparison with our model. RR acknowledges the financial support of the Science and Technology Facilities Council via a postgraduate studentship.

\bibliographystyle{mn2e}
\bibliography{Rollins2014V3}

\begin{thebibliography}{35}
\expandafter\ifx\csname natexlab\endcsname\relax\def\natexlab#1{#1}\fi

\bibitem[{Arce {et~al}\mbox{.}(2013)Arce, Mardones, Corder, Garay,
  Noriega-Crespo, \& Raga}]{Aea:2013}
Arce H.~G., Mardones D., Corder S.~A., Garay G., Noriega-Crespo A., Raga A.~C.,
  2013, ApJ, 774, 39

\bibitem[{Bachiller \& Perez-Gutierrez(1997)}]{BP:1997}
Bachiller R., Perez-Gutierrez M., 1997, ApJ, 487, L93

\bibitem[{Bakes \& Tielens(1998)}]{BT:1998}
Bakes E. L.~O., Tielens A. G. G.~M., 1998, ApJ, 499, 258

\bibitem[{Bate, Tricco \& Price(2014)Bate, Tricco, \& Price}]{BTP:2014}
Bate M.~R., Tricco T.~S., Price D.~J., 2014, MNRAS, 437, 77

\bibitem[{Bruderer {et~al}\mbox{.}(2009)Bruderer, Benz, Doty, van Dishoeck, \&
  Bourke}]{Bea:2009}
Bruderer S., Benz A.~O., Doty S.~D., van Dishoeck E.~F., Bourke T.~L., 2009,
  ApJ, 700, 872

\bibitem[{B\"{u}rzle {et~al}\mbox{.}(2011)B\"{u}rzle, Clark, Stasyszyn,
  k~Dolag, \& Klessen}]{Bea:2011}
B\"{u}rzle F., Clark P.~C., Stasyszyn F., k~Dolag, Klessen S., 2011, MNRAS,
  417, L61

\bibitem[{Cant\'{o}, Raga \& Williams(2008)Cant\'{o}, Raga, \&
  Williams}]{CRW:2008}
Cant\'{o} J., Raga A.~C., Williams D.~A., 2008, Revista Mexicana de
  Astronom\'{i}a y Astrof\'{i}sica, 44(2), 293

\bibitem[{Codella {et~al}\mbox{.}(2006)Codella, Viti, Williams, \&
  Bachiller}]{CVWB:2006}
Codella C., Viti S., Williams D.~A., Bachiller R., 2006, ApJ, 644, L41

\bibitem[{Cunningham {et~al}\mbox{.}(2006)Cunningham, Frank, Quillen, \&
  Blackman}]{Cea:2006}
Cunningham A.~J., Frank A., Quillen A.~C., Blackman E.~G., 2006, ApJ, 653, 416

\bibitem[{Gueth, Guilloteau \& Bachiller(1996)Gueth, Guilloteau, \&
  Bachiller}]{GGB:1996}
Gueth F., Guilloteau S., Bachiller R., 1996, A\&A, 307, 891

\bibitem[{Hara {et~al}\mbox{.}(2013)Hara, Shimajiri, Tsukagoshi, Kurono, Saigo,
  Nakamura, Saito, Wilner, \& Kawabe}]{Hea:2013}
Hara C. {et~al.}, 2013, ApJ, 771, 128

\bibitem[{Herczeg {et~al}\mbox{.}(2012)Herczeg, Karska, Bruderer, Kristensen,
  van Dishoeck, Jørgensen, Visser, Wampfler, Bergin, Yıldız, Pontoppidan, \&
  Gracia-Carpio}]{Hea:2012}
Herczeg G.~J. {et~al.}, 2012, A\&A, 540, 84

\bibitem[{Hogerheijde {et~al}\mbox{.}(1998)Hogerheijde, van Dishoeck, Blake, \&
  van Langevelde}]{Hea:1998}
Hogerheijde M.~R., van Dishoeck E.~F., Blake G.~A., van Langevelde H.~J., 1998,
  ApJ, 502, 315

\bibitem[{Jim\'{e}nez-Serra {et~al}\mbox{.}(2006)Jim\'{e}nez-Serra,
  Martin-Pintado, Viti, Martin, Rodriguez-Franco, Faure, \&
  Tennyson}]{JSea:2006}
Jim\'{e}nez-Serra I., Martin-Pintado J., Viti S., Martin S., Rodriguez-Franco
  A., Faure A., Tennyson J., 2006, ApJ, 605, L135

\bibitem[{Kama {et~al}\mbox{.}(2013)Kama, L\'{o}pez-Sepulcre, Dominik,
  Ceccarelli, Fuente, Caux, Higgins, Tielens, \& Alonso-Albi}]{Kea:2013}
Kama M. {et~al.}, 2013, A\&A, 556, 57

\bibitem[{Kristensen {et~al}\mbox{.}(2010)Kristensen, Visser, van Dishoeck,
  Y{\i}ld{\i}z, Doty, Herczeg, Liu, Parise, Jørgensen, van Kempen, Brinch,
  Wampfler, \& Bruderer}]{Kea:2010}
Kristensen L.~E. {et~al.}, 2010, A\&A, 521, L30

\bibitem[{Langer, Velusamy \& Xie(1996)Langer, Velusamy, \& Xie}]{LVX:1996}
Langer W.~D., Velusamy T., Xie T., 1996, ApJ, 468, L41

\bibitem[{Li {et~al}\mbox{.}(2013)Li, Qiu, Wyrowski, \& Menten}]{LWM:2013}
Li G., Qiu K., Wyrowski F., Menten K., 2013, A\&A, 559, L23

\bibitem[{Montmerle(2001)}]{M:2001}
Montmerle T., 2001, ASPC, 243, 731

\bibitem[{Nisini {et~al}\mbox{.}(2010)Nisini, Benedettini, Codella, Giannini,
  Liseau, Neufeld, Tafalla, van Dishoeck, Bachiller, Baudry, Benz, Bergin,
  Bjerkeli, Blake, Bontemps, Braine, Bruderer, Caselli, Cernicharo, \&
  Daniel}]{Nea:2010}
Nisini B. {et~al.}, 2010, A\&A, 518, L120

\bibitem[{Qiu {et~al}\mbox{.}(2009)Qiu, Zhang, Wu, \& Chen}]{Qea:2009}
Qiu K., Zhang Q., Wu J., Chen H., 2009, ApJ, 696, 66

\bibitem[{Rawlings, Redman \& Carolan(2013)Rawlings, Redman, \&
  Carolan}]{Rea:2013}
Rawlings J. M.~C., Redman M.~P., Carolan P.~B., 2013, MNRAS, 435, 289

\bibitem[{Rawlings {et~al}\mbox{.}(2004)Rawlings, Redman, Keto, \&
  Williams}]{Rea:2004}
Rawlings J. M.~C., Redman M.~P., Keto E., Williams D.~A., 2004, MNRAS, 351,
  1054

\bibitem[{Rawlings, Taylor \& Williams(2000)Rawlings, Taylor, \&
  Williams}]{RTW:2000}
Rawlings J. M.~C., Taylor S.~D., Williams D.~A., 2000, MNRAS, 313, 461

\bibitem[{Schoeier {et~al}\mbox{.}(2005)Schoeier, van~der Tak, van Dishoeck, \&
  Black}]{Sea:2005}
Schoeier F.~L., van~der Tak F. F.~S., van Dishoeck E.~F., Black J.~H., 2005,
  A\&A, 432, 369

\bibitem[{Shu(1977)}]{S:1977}
Shu F., 1977, ApJ, 214, 488

\bibitem[{Spaans {et~al}\mbox{.}(1995)Spaans, Hogerheijde, Mundy, \& van
  Dishoeck}]{Sea:1995}
Spaans M., Hogerheijde M.~R., Mundy L.~G., van Dishoeck E.~F., 1995, ApJ, 455,
  L167

\bibitem[{St\"{a}uber {et~al}\mbox{.}(2005)St\"{a}uber, Doty, van Dishoeck, \&
  Benz}]{SDvDB:2005}
St\"{a}uber P., Doty S.~D., van Dishoeck E.~F., Benz A.~O., 2005, A\&A, 440,
  949

\bibitem[{van~der Tak {et~al}\mbox{.}(2007)van~der Tak, Black, Sch\"{o}ier,
  Jansen, \& van Dishoeck}]{vdTea:2007}
van~der Tak F., Black J.~H., Sch\"{o}ier F.~L., Jansen D.~J., van Dishoeck
  E.~F., 2007, A\&A, 468, 627

\bibitem[{van Kempen {et~al}\mbox{.}(2010)van Kempen, Kristensen, Herczeg,
  Visser, van Dishoeck, Wampfler, Bruderer, Benz, Doty, Brinch, Hogerheijde,
  Jørgensen, Tafalla, Neufeld, Bachiller, Baudry, Benedettini, \&
  Bergin}]{vKea:2010}
van Kempen T.~A. {et~al.}, 2010, A\&A, 518, L121

\bibitem[{van Kempen {et~al}\mbox{.}(2009)van Kempen, van Dishoeck, G\"{u}sten,
  Kristensen, Schilke, Hogerheijde, Boland, Nefs, Menten, \&
  Wyrowski}]{vKea:2009}
van Kempen T.~A. {et~al.}, 2009, A\&A, 501, 633

\bibitem[{Visser {et~al}\mbox{.}(2012)Visser, Kristensen, Bruderer, van
  Dishoeck, Herczeg, Brinch, Doty, Harsono, \& Wolfire}]{Vea:2012}
Visser R. {et~al.}, 2012, A\&A, 537, 55

\bibitem[{Woodall {et~al}\mbox{.}(2007)Woodall, Ag\'{u}ndez, Markwick-Kemper,
  \& Millar}]{Wea:2007}
Woodall J., Ag\'{u}ndez M., Markwick-Kemper A.~J., Millar T.~J., 2007, A\&A,
  466, 1197

\bibitem[{Y{\i}ld{\i}z {et~al}\mbox{.}(2012)Y{\i}ld{\i}z, Kristensen, van
  Dishoeck, Belloche, van Kempen, Hogerheijde, Güsten, \& van~der
  Mare}]{Yea:2012}
Y{\i}ld{\i}z U.~A., Kristensen L.~E., van Dishoeck E.~F., Belloche A., van
  Kempen T.~A., Hogerheijde M.~R., Güsten R., van~der Mare N., 2012, A\&A,
  542, 86

\bibitem[{Y{\i}ld{\i}z {et~al}\mbox{.}(2010)Y{\i}ld{\i}z, van Dishoeck,
  Kristensen, Visser, Jørgensen, Herczeg, van Kempen, Hogerheijde, Doty, Benz,
  Bruderer, \& Wampfler}]{Yea:2010}
Y{\i}ld{\i}z U.~A. {et~al.}, 2010, A\&A, 521, L40

\end{thebibliography}

\appendix

\section{Deriving $\lowercase{\psi}$}
\label{app:psi}

As presented in \cite{CRW:2008}, given an arbitrary scale length $r_{\rm 0}$, the dimensionless shape equation $\bar{R}_{\rm S}(\theta)={R_{\rm S}}/{r_0}$ for a momentum driven wind of constant mass loss rate, steadily decollimating at a constant angular rate and expanding into an isothermal sphere, can be defined implicitly as:

\begin{align}
\label{eq:shape}
\bar{\epsilon}^2\bar{\kappa}\bar{R}_{\rm S}^2 + 4(\bar{\epsilon}\bar{R}_{\rm S}&+\theta-\theta_{\rm m})\mathrm{cot}(\theta/2) \nonumber \\
&\qquad+ 8\ln\left[\frac{\sin[(\theta_{\rm m}-\bar{\epsilon}\bar{R}_{\rm S})/2]}{\sin(\theta/2)}\right] = 0
\end{align}

\noindent where $\bar{\epsilon}$ and $\bar{\kappa}$ are the dimensionless outflow decollimation rate and isothermal coefficient respectively and $\theta_{\rm m}$ is the cavity opening angle. This equation can be differentiated with respect to $\theta$ to give an implicit equation for $\bar{R}_{\rm S}' = {{\rm d}\bar{R}_{\rm S}}/{{\rm d}\theta}$:

\begin{align}
\label{eq:diffshape}
2\bar{\epsilon}^2\bar{\kappa}\bar{R}_{\rm S}\bar{R}_{\rm S}'& \nonumber \\
\qquad+ 4(\bar{\epsilon}\bar{R}_{\rm S}'+&1)\mathrm{cot}(\theta/2) \nonumber \\
\qquad- 2(\bar{\epsilon}\bar{R}_{\rm S} &+ \theta - \theta_{\rm m})\mathrm{cosec}^2(\theta/2) \nonumber \\
\qquad- 4\bigg[&\bar{\epsilon}\bar{R}_{\rm S}'\mathrm{cot}\left(\frac{\theta_{\rm m}-\bar{\epsilon}\bar{R}_{\rm S}}{2}\right) + \mathrm{cot}(\theta/2)\bigg] = 0.
\end{align}

\noindent Converting back to dimensional quantities and considering the geometry of the problem in cylindrical coordinates (r,$\phi$,z), we can calculate $\omega$, the angle between the cavity wall tangent and the cylindrical radial vector (r) as:

\begin{equation}
{\rm d}z = {\rm d}R_{\rm S}\cos\theta - R_{\rm S}\sin\theta {\rm d}\theta
\end{equation}
\begin{equation}
{\rm d}r = {\rm d}R_{\rm S}\sin\theta + R_{\rm S}\cos\theta {\rm d}\theta
\end{equation}
\begin{equation}
\label{eq:diffcoords}
\tan\omega = \frac{{\rm d}z}{{\rm d}r} = \frac{R_{\rm S}' - R_{\rm S}\tan\theta}{R_{\rm S}'\tan\theta + R_{\rm S}}.
\end{equation}

\noindent From triangle geometry we then get a function for the mixing angle, $\psi$, as a function of $R_{\rm S},\ R_{\rm S}'$ and $\theta$:

\begin{equation}
\label{eq:psimixderive}
\psi = \mathrm{arctan}\left(\frac{R_{\rm S}' - R_{\rm S}\tan\theta}{R_{\rm S}'\tan\theta + R_{\rm S}}\right) + \theta - \frac{\pi}{2}
\end{equation}

\section{Deriving $\lowercase{\dot{n}_{\rm inj}}$}
\label{app:ndot}

From the definitions given in \cite{CRW:2008}, we define the density injection rate at time $t_{\rm dyn}$ into the interface, $\dot{\rho}_{\rm inj}$, as the mass loss rate per unit solid angle ($\dot{m}_{\rm \Omega_{\rm 0}}$) at a time $R_{\rm S}/v_{\rm 0}$ in the past divided into a box of volume $R_{\rm S}^2l$ where $l = 0.1R_{\rm S}$ is the assumed thickness of the cavity:

\begin{equation}
\label{eq:inj1}
\dot{\rho}_{\rm inj} = \frac{\dot{m}_{\rm \Omega_{\rm 0}}}{R_{\rm S}^2l}\sin\psi
\end{equation}

\noindent Here we have added a nominal mixing efficiency $\sin\psi$ depending on the angle of impact between the outflowing material and the cavity wall, $\psi$, as derived in Appendix \ref{app:psi}. The definitions of these variables from \cite{CRW:2008} then give our equation for the rate of injected particle density, $\dot{n}_{\rm inj}$, assuming it is dominated by H$^+$ ions of mass $m_{\rm H}$:

\begin{equation}
\label{eq:inj2}
\dot{n}_{\rm inj} = \frac{5\dot{M}_{\rm 0}}{\pi m_{\rm H} R_{\rm S}(\theta)^3} \frac{\sin\psi}{1-\cos[\epsilon(t_{\rm dyn}-\frac{R_{\rm S}(\theta)}{v_{\rm 0}})]} \, {\rm cm}^{-3} \, {\rm s}^{-1}
\end{equation}

\noindent where $\dot{M}_{\rm 0}$ is the total mass loss rate of the outflow, $m_{\rm H}$ is the hydrogen atomic mass, $\epsilon$ is the outflow decollimation rate and $v_{\rm 0}$ is the outflow velocity.

\end{document}